\documentclass[ aps,prl,superscriptaddress]{revtex4-1}

\usepackage{color}
\usepackage{graphicx}
\usepackage{bm}
\begin{document}


\title{Evaluation of the Interplanetary Magnetic Field Strength Using the Cosmic-Ray Shadow of the Sun}


\author{M.~Amenomori}
\affiliation{Department of Physics, Hirosaki University, Hirosaki 036-8561, Japan }
\author{X.~J.~Bi}
\affiliation{Key Laboratory of Particle Astrophysics, Institute of High Energy Physics, Chinese Academy of Sciences, Beijing 100049, China }
\author{D.~Chen}
\affiliation{National Astronomical Observatories, Chinese Academy of Sciences, Beijing 100012, China }
\author{T.~L.~Chen}
\affiliation{Department of Mathematics and Physics, Tibet University, Lhasa 850000, China }
\author{W.~Y.~Chen}
\affiliation{Key Laboratory of Particle Astrophysics, Institute of High Energy Physics, Chinese Academy of Sciences, Beijing 100049, China }
\author{S.~W.~Cui}
\affiliation{Department of Physics, Hebei Normal University, Shijiazhuang 050016, China }
\author{Danzengluobu}
\affiliation{Department of Mathematics and Physics, Tibet University, Lhasa 850000, China }
\author{L.~K.~Ding}
\affiliation{Key Laboratory of Particle Astrophysics, Institute of High Energy Physics, Chinese Academy of Sciences, Beijing 100049, China }
\author{C.~F.~Feng}
\affiliation{Department of Physics, Shandong University, Jinan 250100, China }
\author{Zhaoyang~Feng}
\affiliation{Key Laboratory of Particle Astrophysics, Institute of High Energy Physics, Chinese Academy of Sciences, Beijing 100049, China }
\author{Z.~Y.~Feng}
\affiliation{Institute of Modern Physics, SouthWest Jiaotong University, Chengdu 610031, China }
\author{Q.~B.~Gou}
\affiliation{Key Laboratory of Particle Astrophysics, Institute of High Energy Physics, Chinese Academy of Sciences, Beijing 100049, China }
\author{Y.~Q.~Guo}
\affiliation{Key Laboratory of Particle Astrophysics, Institute of High Energy Physics, Chinese Academy of Sciences, Beijing 100049, China }
\author{H.~H.~He}
\affiliation{Key Laboratory of Particle Astrophysics, Institute of High Energy Physics, Chinese Academy of Sciences, Beijing 100049, China }
\author{Z.~T.~He}
\affiliation{Department of Physics, Hebei Normal University, Shijiazhuang 050016, China }
\author{K.~Hibino}
\affiliation{Faculty of Engineering, Kanagawa University, Yokohama 221-8686, Japan }
\author{N.~Hotta}
\affiliation{Faculty of Education, Utsunomiya University, Utsunomiya 321-8505, Japan }
\author{Haibing~Hu}
\affiliation{Department of Mathematics and Physics, Tibet University, Lhasa 850000, China }
\author{H.~B.~Hu}
\affiliation{Key Laboratory of Particle Astrophysics, Institute of High Energy Physics, Chinese Academy of Sciences, Beijing 100049, China }
\author{J.~Huang}
\affiliation{Key Laboratory of Particle Astrophysics, Institute of High Energy Physics, Chinese Academy of Sciences, Beijing 100049, China }
\author{H.~Y.~Jia}
\affiliation{Institute of Modern Physics, SouthWest Jiaotong University, Chengdu 610031, China }
\author{L.~Jiang}
\affiliation{Key Laboratory of Particle Astrophysics, Institute of High Energy Physics, Chinese Academy of Sciences, Beijing 100049, China }
\author{F.~Kajino}
\affiliation{Department of Physics, Konan University, Kobe 658-8501, Japan }
\author{K.~Kasahara}
\affiliation{Research Institute for Science and Engineering, Waseda University, Tokyo 169-8555, Japan }
\author{Y.~Katayose}
\affiliation{Faculty of Engineering, Yokohama National University, Yokohama 240-8501, Japan }
\author{C.~Kato}
\affiliation{Department of Physics, Shinshu University, Matsumoto 390-8621, Japan }
\author{K.~Kawata}
\affiliation{Institute for Cosmic Ray Research, University of Tokyo, Kashiwa 277-8582, Japan }
\author{M.~Kozai}
\affiliation{Department of Physics, Shinshu University, Matsumoto 390-8621, Japan }
\affiliation{Institute of Space and Astronautical Science, Japan Aerospace Exploration Agency (ISAS/JAXA), Sagamihara, Kanagawa 252-5210, Japan }
\author{Labaciren}
\affiliation{Department of Mathematics and Physics, Tibet University, Lhasa 850000, China }
\author{G.~M.~Le}
\affiliation{National Center for Space Weather, China Meteorological Administration, Beijing 100081, China }
\author{A.~F.~Li}
\affiliation{School of Information Science and Engineering, Shandong Agriculture University, Taian 271018, China }
\affiliation{Department of Physics, Shandong University, Jinan 250100, China }
\affiliation{Key Laboratory of Particle Astrophysics, Institute of High Energy Physics, Chinese Academy of Sciences, Beijing 100049, China }
\author{H.~J.~Li}
\affiliation{Department of Mathematics and Physics, Tibet University, Lhasa 850000, China }
\author{W.~J.~Li}
\affiliation{Key Laboratory of Particle Astrophysics, Institute of High Energy Physics, Chinese Academy of Sciences, Beijing 100049, China }
\affiliation{Institute of Modern Physics, SouthWest Jiaotong University, Chengdu 610031, China }
\author{C.~Liu}
\affiliation{Key Laboratory of Particle Astrophysics, Institute of High Energy Physics, Chinese Academy of Sciences, Beijing 100049, China }
\author{J.~S.~Liu}
\affiliation{Key Laboratory of Particle Astrophysics, Institute of High Energy Physics, Chinese Academy of Sciences, Beijing 100049, China }
\author{M.~Y.~Liu}
\affiliation{Department of Mathematics and Physics, Tibet University, Lhasa 850000, China }
\author{H.~Lu}
\affiliation{Key Laboratory of Particle Astrophysics, Institute of High Energy Physics, Chinese Academy of Sciences, Beijing 100049, China }
\author{X.~R.~Meng}
\affiliation{Department of Mathematics and Physics, Tibet University, Lhasa 850000, China }
\author{T.~Miyazaki}
\affiliation{Department of Physics, Shinshu University, Matsumoto 390-8621, Japan }
\author{K.~Mizutani\footnote[2]{Deceased.}}
\affiliation{Research Institute for Science and Engineering, Waseda University, Tokyo 169-8555, Japan }
\affiliation{Saitama University, Saitama 338-8570, Japan }
\author{K.~Munakata}
\affiliation{Department of Physics, Shinshu University, Matsumoto 390-8621, Japan }
\author{T.~Nakajima}
\affiliation{Department of Physics, Shinshu University, Matsumoto 390-8621, Japan }
\author{Y.~Nakamura}
\email[]{15st303c@shinshu-u.ac.jp}
\affiliation{Department of Physics, Shinshu University, Matsumoto 390-8621, Japan }
\author{H.~Nanjo}
\affiliation{Department of Physics, Hirosaki University, Hirosaki 036-8561, Japan }
\author{M.~Nishizawa}
\affiliation{National Institute of Informatics, Tokyo 101-8430, Japan }
\author{T.~Niwa}
\affiliation{Department of Physics, Shinshu University, Matsumoto 390-8621, Japan }
\author{M.~Ohnishi}
\affiliation{Institute for Cosmic Ray Research, University of Tokyo, Kashiwa 277-8582, Japan }
\author{I.~Ohta}
\affiliation{Sakushin Gakuin University, Utsunomiya 321-3295, Japan }
\author{S.~Ozawa}
\affiliation{Research Institute for Science and Engineering, Waseda University, Tokyo 169-8555, Japan }
\author{X.~L.~Qian}
\affiliation{Department of Physics, Shandong University, Jinan 250100, China }
\affiliation{Key Laboratory of Particle Astrophysics, Institute of High Energy Physics, Chinese Academy of Sciences, Beijing 100049, China }
\author{X.~B.~Qu}
\affiliation{College of Science, China University of Petroleum, Qingdao 266555, China }
\author{T.~Saito}
\affiliation{Tokyo Metropolitan College of Industrial Technology, Tokyo 116-8523, Japan }
\author{T.~Y.~Saito}
\affiliation{Max-Planck-Institut f\"ur Physik, M\"unchen D-80805, Deutschland }
\author{M.~Sakata}
\affiliation{Department of Physics, Konan University, Kobe 658-8501, Japan }
\author{T.~K.~Sako}
\affiliation{Institute for Cosmic Ray Research, University of Tokyo, Kashiwa 277-8582, Japan }
\affiliation{Escuela de Ciencias F\'{\i}sicas y Nanotechnolog\'{\i}a, Yachay Tech, Imbabura 100115, Ecuador }
\author{J.~Shao}
\affiliation{Key Laboratory of Particle Astrophysics, Institute of High Energy Physics, Chinese Academy of Sciences, Beijing 100049, China }
\affiliation{Department of Physics, Shandong University, Jinan 250100, China }
\author{M.~Shibata}
\affiliation{Faculty of Engineering, Yokohama National University, Yokohama 240-8501, Japan }
\author{A.~Shiomi}
\affiliation{College of Industrial Technology, Nihon University, Narashino 275-8576, Japan }
\author{T.~Shirai}
\affiliation{Faculty of Engineering, Kanagawa University, Yokohama 221-8686, Japan }
\author{H.~Sugimoto}
\affiliation{Shonan Institute of Technology, Fujisawa 251-8511, Japan }
\author{M.~Takita}
\affiliation{Institute for Cosmic Ray Research, University of Tokyo, Kashiwa 277-8582, Japan }
\author{Y.~H.~Tan}
\affiliation{Key Laboratory of Particle Astrophysics, Institute of High Energy Physics, Chinese Academy of Sciences, Beijing 100049, China }
\author{N.~Tateyama}
\affiliation{Faculty of Engineering, Kanagawa University, Yokohama 221-8686, Japan }
\author{S.~Torii}
\affiliation{Research Institute for Science and Engineering, Waseda University, Tokyo 169-8555, Japan }
\author{H.~Tsuchiya}
\affiliation{Japan Atomic Energy Agency, Tokai-mura 319-1195, Japan }
\author{S.~Udo}
\affiliation{Faculty of Engineering, Kanagawa University, Yokohama 221-8686, Japan }
\author{H.~Wang}
\affiliation{Key Laboratory of Particle Astrophysics, Institute of High Energy Physics, Chinese Academy of Sciences, Beijing 100049, China }
\author{H.~R.~Wu}
\affiliation{Key Laboratory of Particle Astrophysics, Institute of High Energy Physics, Chinese Academy of Sciences, Beijing 100049, China }
\author{L.~Xue}
\affiliation{Department of Physics, Shandong University, Jinan 250100, China }
\author{Y.~Yamamoto}
\affiliation{Department of Physics, Konan University, Kobe 658-8501, Japan }
\author{K.~Yamauchi}
\affiliation{Faculty of Engineering, Yokohama National University, Yokohama 240-8501, Japan }
\author{Z.~Yang}
\affiliation{Key Laboratory of Particle Astrophysics, Institute of High Energy Physics, Chinese Academy of Sciences, Beijing 100049, China }
\author{A.~F.~Yuan}
\affiliation{Department of Mathematics and Physics, Tibet University, Lhasa 850000, China }
\author{T.~Yuda$^{\rm b}$}
\affiliation{Institute for Cosmic Ray Research, University of Tokyo, Kashiwa 277-8582, Japan }
\author{L.~M.~Zhai}
\affiliation{National Astronomical Observatories, Chinese Academy of Sciences, Beijing 100012, China }
\author{H.~M.~Zhang}
\affiliation{Key Laboratory of Particle Astrophysics, Institute of High Energy Physics, Chinese Academy of Sciences, Beijing 100049, China }
\author{J.~L.~Zhang}
\affiliation{Key Laboratory of Particle Astrophysics, Institute of High Energy Physics, Chinese Academy of Sciences, Beijing 100049, China }
\author{X.~Y.~Zhang}
\affiliation{Department of Physics, Shandong University, Jinan 250100, China }
\author{Y.~Zhang}
\affiliation{Key Laboratory of Particle Astrophysics, Institute of High Energy Physics, Chinese Academy of Sciences, Beijing 100049, China }
\author{Yi~Zhang}
\affiliation{Key Laboratory of Particle Astrophysics, Institute of High Energy Physics, Chinese Academy of Sciences, Beijing 100049, China }
\author{Ying~Zhang}
\affiliation{Key Laboratory of Particle Astrophysics, Institute of High Energy Physics, Chinese Academy of Sciences, Beijing 100049, China }
\author{Zhaxisangzhu}
\affiliation{Department of Mathematics and Physics, Tibet University, Lhasa 850000, China }
\author{X.~X.~Zhou}
\affiliation{Institute of Modern Physics, SouthWest Jiaotong University, Chengdu 610031, China }


\collaboration{The Tibet AS$\gamma$ Collaboration}
\noaffiliation

\date{\today}

\begin{abstract}
We analyze the Sun's shadow observed with the Tibet-III air shower array and find that the shadow's center deviates northward (southward) from the optical solar disc center in the ``Away'' (``Toward'') IMF sector. 
By comparing with numerical simulations based on the solar magnetic field model, we find that the average IMF strength in the ``Away'' (``Toward'') sector is $1.54 \pm 0.21_{\rm stat} \pm 0.20_{\rm syst}$ ($1.62 \pm 0.15_{\rm stat} \pm 0.22_{\rm syst}$) times larger than the model prediction. 
These demonstrate that the observed Sun's shadow is a useful tool for the quantitative evaluation of the average solar magnetic field.

\end{abstract}


\maketitle

\section{\label{sec:s1}Introduction}
The Sun blocks cosmic rays arriving at the Earth from the direction of the Sun and casts a shadow in the cosmic-ray intensity. 
Cosmic rays are positively charged particles, consisting of mostly protons and helium nuclei, and their trajectories are deflected by the magnetic field between the Sun and Earth, depending on the magnetic field strength $B$ and polarity, and on the cosmic ray rigidity. 
The Tibet air shower (AS) experiment has successfully observed the Sun's shadow at 10 TeV energies and has confirmed, for the first time, the small but the measurable effect of the solar magnetic field on the shadow \cite{Amenomori2013}. 
The observed intensity deficit in the Sun's shadow shows a clear 11-year variation decreasing with increasing solar activity.
Our numerical simulations succeeded in reproducing this observed feature quantitatively and showed that, during solar maximum, cosmic rays passing near the solar limb are ``scattered'' by the strong coronal magnetic field and may appear from the direction of the optical solar disc and reduce the intensity deficit of the Sun's shadow.\par

While the strong coronal magnetic field affects the intensity deficit in the shadow, the interplanetary magnetic field (IMF) between the Sun and Earth also deflects orbits of TeV cosmic rays. 
This deflection has actually been observed by the AS experiments as a North-South displacement of the center of the Sun's shadow from the optical center of the Sun \cite{Amenomori2000,ARGO}.\par

These observations indicate that the Sun's shadow can be used as a sensor of the solar magnetic field. 
The solar magnetic field on the photosphere has been continuously monitored by optical measurements using the Zeeman effect \cite{Jones1992}, while the local IMF at the Earth has been directly observed by the near Earth satellites \cite{OMNI_web}. 
The observation of the average IMF between the Sun and Earth, however, still remains difficult. 
Since the orbital deflection of cosmic rays is proportional to $B$, the observed Sun's shadow can be used for evaluating the large-scale IMF averaged between the Sun and Earth.\par

In this Letter, we analyze the angular displacement of the shadow's center observed by the Tibet AS array and evaluate the IMF strength $B_{\rm IMF}$ by comparing the observation with detailed numerical simulations based on the potential field model (PFM) of the solar magnetic field, which describes the IMF in terms of the observed photospheric magnetic field. 
We shall demonstrate that average $\overline{B_{\rm IMF}}$ is significantly underestimated by the widely used PFM \cite{Wiegelmann2015}.

\section{\label{sec:s2}Experiment and data analysis}

We analyze the Sun's shadow observed between 2000 March and 2009 August by the Tibet-III AS array which has been operating since late 1999 at Yangbajing (4,300 m above sea level) in Tibet, China. 
The Tibet-III AS array consists of 789 scintillation detectors with a 7.5 m spacing, each with 0.5 m$^2$ detection area, covering an effective area of 37,000 m$^2$ 
\cite{Amenomori2009}. 
In this paper, we divide the observed AS events into seven energy bins according to their shower size $\sum \rho_{\rm FT}$, which is the sum of the number of particles per m$^2$ for each fast-timing (FT) detector and used as a measure of the primary cosmic-ray energy. 
For $\sum\rho_{\rm FT}$ we consider the intervals:$17.8<\sum\rho_{\rm FT}\leq31.6$, $31.6<\sum\rho_{\rm FT}\leq56.2$, $56.2<\sum\rho_{\rm FT}\leq100$, $100<\sum\rho_{\rm FT}\leq215$, $215<\sum\rho_{\rm FT}\leq464$, $464<\sum\rho_{\rm FT}\leq1000$, and $\sum\rho_{\rm FT}>1000$. 
The modal energies of primary cosmic rays corresponding to these energy bins are 4.9, 7.7, 13, 22, 43, 90 and 240 TeV, respectively, and the ``window size'' $\bigtriangleup d$, which is angular distance from true direction including 68\% events estimated by MC simulation, are $2.0^{\circ}, 1.4^{\circ}, 0.9^{\circ}, 0.6^{\circ}, 0.4^{\circ}, 0.3^{\circ}$, and $0.2^{\circ}$, respectively. 
These modal energies of Tibet-III extending below $\sim$10 TeV are suitable for analyzing the angular displacement of the Sun's shadow, because the magnetic deflection is expected to be larger for lower energy cosmic rays.

 \begin{figure}
 \includegraphics[width=8.5cm]{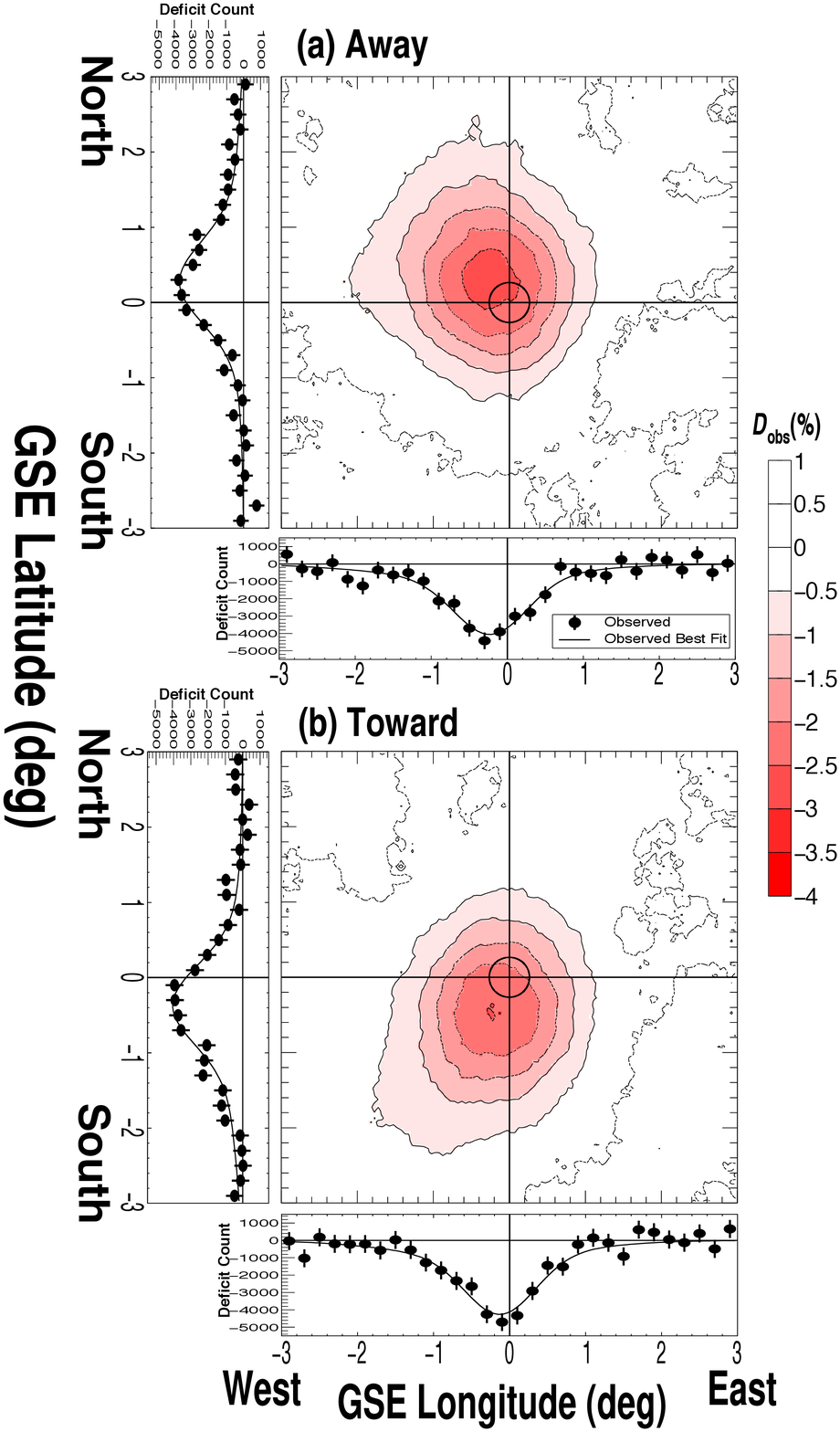}
 \caption{\label{fig:f1} Two dimensional maps of $D_{\rm{obs}}$ in ``Away'' (top) and ``Toward'' (bottom) sectors in 2000-2009. 
 Each panel shows two dimensional contour of $D_{\rm{ obs}}$ deduced from AS events with $\sum \rho_{\mathrm{FT}}>10$ corresponding to the modal primary energy of $\sim3$ TeV and $\bigtriangleup d=0.9^{\circ}$. 
 In each panel, a small circle centered on the origin indicates the optical solar disc. 
 The significance at maximum deficit point derived with Li-Ma formula in ``Away'' (``Toward'') sector is -23.9$\sigma$ (-22.1$\sigma$) \cite{LiMa1983}.
 The projections of $D_{\rm obs}$ on the horizontal and vertical axes are also attached to each panel.
}
 \end{figure}

For the analysis of the Sun's shadow, the number of on-source events ($N_{\rm on}$) is defined as the number of AS events arriving from the direction within a circle of $\bigtriangleup d$ radius centered at a given point on the celestial sphere. 
The number of background or off-source events ($\langle N_{\rm off}\rangle$) is calculated by averaging the number of events within each of the eight off-source windows which are located at the same zenith angle as the on-source window \cite{Amenomori2013}. 
We then estimate the intensity deficit relative to the number of background events as $D_{\rm obs}=(N_{\rm on}-\langle N_{\rm off}\rangle)/\langle N_{\rm off}\rangle$ at every $0.1^{\circ}$ grid of the Geocentric Solar Ecliptic (GSE) longitude and latitude surrounding the optical center of the Sun toward which the GSE-X axis directs from the Earth. 
We confirmed a clear 11-year variation of $D_{\rm obs}$ being successfully observed also by Tibet-III AS array. 
We will report this elsewhere.\par
We assign the IMF sector polarity to each day referring to the daily mean GSE-$x$ and GSE-$y$ components of the IMF ($B_x$, $B_y$) observed by near Earth satellites \cite{OMNI_web} and calculate $D_{\rm obs}$ in ``Away'' and ``Toward'' sectors, separately. 
We assign ``Away'' (``Toward'') sector polarity to a day when the IMF observed two days later satisfies $B_x<0$ and $B_y>0$ ($B_x>0$ and $B_y<0$) and ``unknown'' to the remaining days. 
The sector polarity in the solar corona is carried out by the solar wind with an average velocity of $\sim$400 km/s and observed at the Earth about four days later. 
For our assignment of the IMF sector polarity to a day under consideration, therefore, we use the IMF data observed at the Earth two days later as an average along the Sun-Earth line on the day. 
In about 65 \% of ``Away'' or ``Toward" days assigned in this way, a different polarity is observed over following four days, indicating the mixed polarity along the Sun-Earth line. We confirmed, however, that basic conclusions obtained below in this paper remain unchanged even by excluding these mixed polarity days from our analyses.\par

\begin{figure*}
 \includegraphics[width=17.7cm]{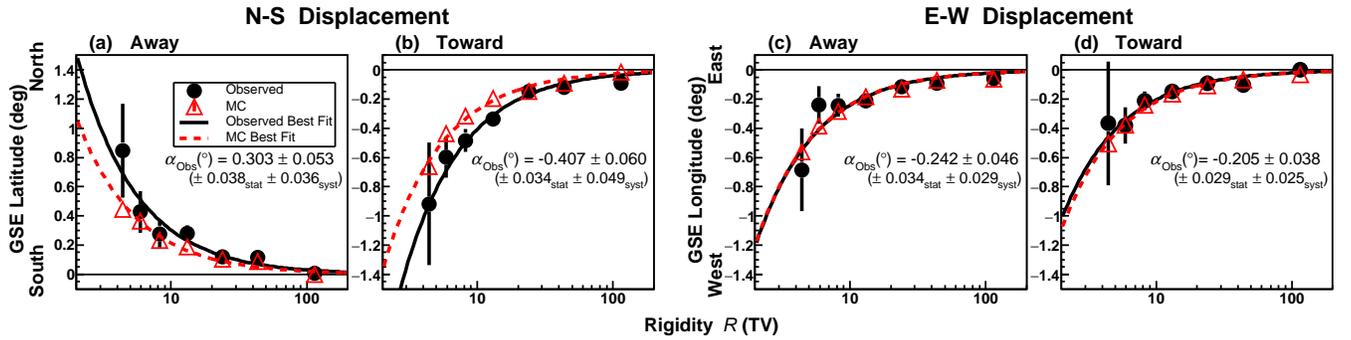}
 \caption{\label{fig:f2} The rigidity dependences of the N-S ((a) and (b)) and E-W ((c) and (d)) displacements of the center of the Sun's shadow in ``Away'' and ``Toward'' sectors. 
 Black solid circles and red open triangles in each panel show the observed and simulated displacements, respectively, each as a function of the rigidity ($R$) on the horizontal axis. 
 The error bar of each solid circle indicates the statistical error. 
 Black solid and red broken curves display the function of $\alpha/(R/ \rm{10TV})$ best-fitting to black solid circles and red open triangles, respectively. 
 The best-fit parameter $\alpha$ to the observed data is indicated in each panel with systematic errors estimated from the systematic error of the primary energy in our analyses of the Moon's shadow \cite{Amenomori2009}.}
 \end{figure*}

Figure \ref{fig:f1} shows $D_{\rm obs}$ in \% deduced from all AS events in ``Away'' and ``Toward'' sectors, each as a function of the GSE latitude and longitude measured from the optical Sun's center, together with each projection on the vertical (North-South: N-S) or horizontal (East-West: E-W) axis. Following the method developed for our analyses of the Moon's shadow \cite{Amenomori2009}, we deduce the angular distance of the shadow's center from the optical Sun's center by best-fitting the model function to the N-S and E-W projections. 
It is seen in Figure \ref{fig:f1} that the shadow's center clearly deviates from the optical center of the Sun at the origin of the map. 
The shadow's center shifts northward (southward) in ``Away'' (``Toward'') sector as expected from the deflection in the average positive (negative) $B_y$ along the Sun-Earth line, while the shadow's center shifts westward regardless of the IMF sector polarity. 
In Figure \ref{fig:f2}, the average N-S and E-W displacement angles in ``Away'' and ``Toward'' sectors are calculated for each energy bin and plotted as functions of $R$ denoting the average rigidity of cosmic rays which are blocked by the Sun. 
We convert the modal energy of each energy bin to $R$ using the energy spectra and elemental composition of primary cosmic rays reported mainly from the direct measurements \cite{Shibata2010}. 
As expected from the magnetic deflection of charged particles, the observed displacement angles displayed by black solid circles are reasonably well fitted by a function $\alpha /(R/10 \mathrm{TV})$ of $R$ in TV with a fitting parameter $\alpha$ denoting the displacement angle at 10 TV. 

\section{\label{sec:s3}MC simulation}
	In order to interpret the observed Sun's shadow, we have carried out detailed Monte Carlo (MC) simulations, tracing orbits of anti-particles shot back from the Earth to the Sun in the model magnetic field between the Sun and Earth \cite{Amenomori2013}.
For the solar magnetic field in the MC simulations, we use the PFM called the current sheet source surface (CSSS) model \cite{ZhaoHoeksema1995}. 
The PFM is unique in the sense that it gives the coronal and interplanetary magnetic field in an integrated manner based on the observed photospheric magnetic field \cite{Wiegelmann2015}. 
The CSSS model involves four free parameters, the radius $R_{\rm ss}$ of the source surface (SS) where the supersonic solar wind stars blowing, the order $n$ of the spherical harmonic series describing the observed photospheric magnetic field, the radius $R_{\rm cp} (<R_{\rm ss})$ of the spherical surface where the magnetic cusp structure in the helmet streamers appears, and the length scale $l_{\rm a}$ of the horizontal electric currents in the corona. 
In our simulations, we set $R_{\rm cp}$ and $l_{\rm a}$ to 1.7 and 1.0 solar radii (1.7$R_{\odot}$ and 1.0$R_{\odot}$), respectively, and $n=10$ which is sufficient to describe the structures relevant to the orbital motion of high-energy particles. 
We also set $R_{\rm ss}$ to $10R_{\odot}$ which gained recent support from observational evidences \cite{SchusslerBaumann2006}. 
Our simulations with this CSSS model reproduces the observed 11-year variation of $D_{\rm obs}$ at 10TeV most successfully \cite{Amenomori2013}. 
The magnetic field components are calculated at each point in the solar corona between $R_{\odot}$ and $R_{\rm ss}$ in terms of the spherical harmonic coefficients derived from the photospheric magnetic field observations with the spectromagnetograph of the National Solar Observatory at Kitt Peak (KPVT/SOLIS) for every Carrington rotation period ($\sim$27.3 days) \cite{Jones1992}. 
We calculate anti-particle orbits by properly rotating the reproduced magnetic field in every Carrington rotation period.
The radial coronal field on the SS is then stretched out to the interplanetary space forming the simple Parker-spiral IMF. 
For the radial solar wind speed needed for the Parker-spiral IMF, we use the ``solar wind speed synoptic chart'' estimated from the interplanetary scintillation measurement in each Carrington rotation and averaged over the Carrington longitude \cite{STE_labWeb}.

 In addition, we assume a stable dipole field for the geomagnetic field \cite{Amenomori2009}.

\section{\label{sec:s4}Results and discussions}
	
 \begin{figure}
 \includegraphics[width=7.5cm]{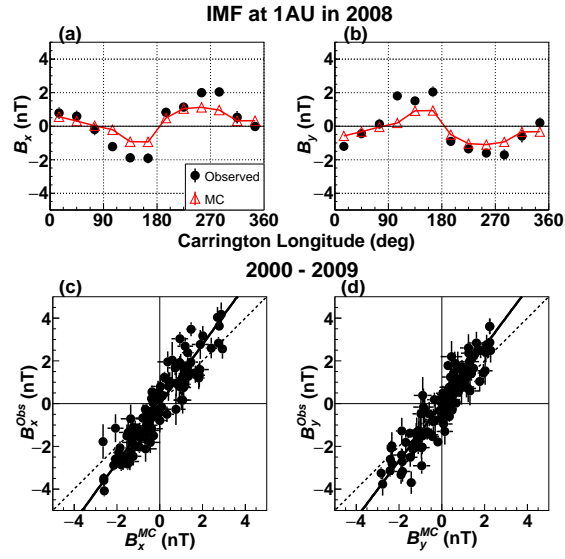}
 \caption{\label{fig:f3} Comparison between the observed and simulated IMF components at the Earth. 
 Black solid circles (red open triangles) in the upper panels display the observed (simulated) $B_x$ (a) and $B_y$ (b), respectively, each as a function of the Carrington longitude. 
 The lower panels show the correlations between the observed and simulated $B_x$ (c) and $B_y$ (d). 
 The regression coefficient (slope of black solid line), is significantly larger than 1 (slope of dashed line). } 
 \end{figure}
 
 \begin{figure}
 \includegraphics[width=7.5cm]{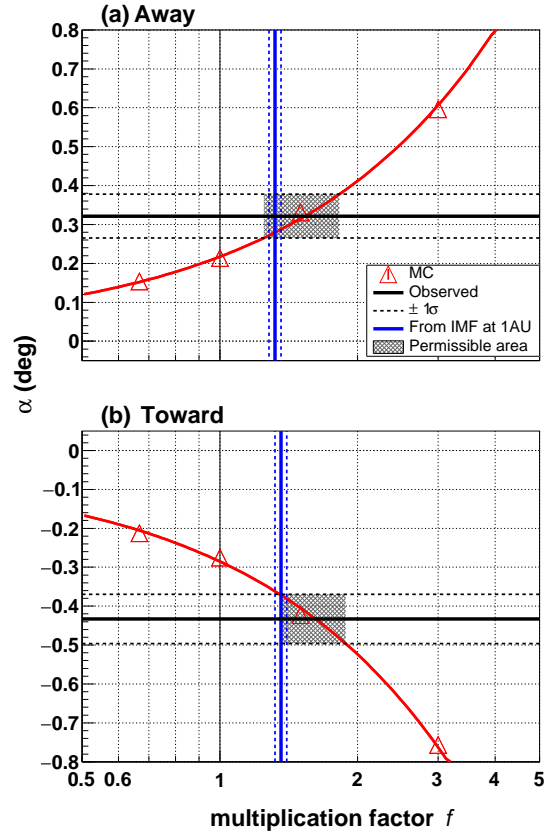}
 \caption{\label{fig:f4}The parameter $\alpha$ obtained from simulations by changing the multiplication factor $f$ (see text). 
 Red open triangles in the top and bottom panels show $\alpha$ in ``Away" and ``Toward" sectors, respectively, each as a function of $f$. 
 The horizontal black line in each panel shows the observed $\alpha$, while the vertical blue line indicates $f$ deduced from the observed and simulated $B_{\rm IMF}$ at 1 AU. 
 The shaded area in each panel indicates one sigma region of $f$ allowed by the observed $\alpha$.}
 \end{figure}

In order to compare observations with predictions, we calculate the N-S and E-W displacements of the simulated shadow's center in ``Away'' and ``Toward'' sectors for various rigidities $R$. 
It is seen in the N-S displacement in Figure \ref{fig:f2}(a) and \ref{fig:f2}(b) that the magnitudes of the simulated displacement (red broken curves) are significantly smaller than the observations (black curves) in both sectors, implying a systematic underestimation by the simulations. 
The simulated E-W displacements in Figure \ref{fig:f2}(c) and \ref{fig:f2}(d), on the other hand, are quite consistent with the observations, implying that the E-W displacements is predominantly arising from the deflection of cosmic ray orbits in the geomagnetic field. 
We confirmed that the E-W displacement of the Sun's shadow is consistent with the Moon's shadow, when an additional minor deflection in the solar magnetic field is taken into account\cite{Amenomori2009}.\par

We also compare the $B_{\rm IMF}$ observed at the Earth with the simulation in Figures \ref{fig:f3}(a) and \ref{fig:f3}(b) and find that the magnitudes of the simulated $B_x$ and $B_y$ are systematically smaller than the observations. 
By calculating the average $B_x$ and $B_y$ each as a function of the Carrington longitude in every year, we examined the correlations between the simulated and observed $B_x$ and $B_y$ as shown in Figures \ref{fig:f3}(c) and \ref{fig:f3}(d). 
While the correlation coefficient between the simulated and observed magnetic field component in this figure is 0.93 (0.92) for $B_x$ ($B_y$), indicating high correlation, the regression coefficient is $1.38 \pm 0.03$ ($1.34 \pm 0.03$ ) for $B_x$ ($B_y$), significantly larger than 1.00, implying the underestimation of the simulated $B_x$ ($B_y$) on the horizontal axes\cite{Jackson2016}.
 This underestimation is observed in every year, while the magnitude of $B_x$ or $B_y$ changes in a positive correlation with the solar activity. 
 We confirmed that the observed average $B_z$ is insignificant in both sectors as expected from the Parker-spiral IMF.\par

 The N-S displacement of the center of Sun's shadow reflects $\overline{B_{\rm IMF}}$ along the cosmic ray orbits between the Sun and Earth, while $B_x$ and $B_y$ in Figure \ref{fig:f3} are the local field components at the Earth. 
 The underestimation of the N-S displacement in Figure \ref{fig:f2}, therefore, inevitably suggests that $\overline{B_{\rm IMF}}$ is underestimated. 
 In order to quantitatively evaluate this underestimation, we simply multiply the simulated $B$ by a constant factor $f$ everywhere in the space outside the geomagnetic field, repeat simulations by changing $f$ and calculate $\alpha(f)$ best-fitting to each simulated displacement. 
 Figure \ref{fig:f4} displays $\alpha(f)$ by red open triangles with linear best-fit curves, each as a function of the multiplication factor $f$. 
 From the intersection between the red curves and black lines showing $\alpha$ for the observed N-S displacement in Figure \ref{fig:f2}, we evaluate $f$ best reproducing the observed displacement to be $1.54 \pm 0.21_{\rm stat} \pm 0.20_{\rm syst}$ $(1.62 \pm 0.15_{\rm stat} \pm 0.22_{\rm syst})$ in ``Away'' (``Toward'') sector. 
 This is consistent with the regression coefficient, $1.38 \pm 0.03$ $(1.34 \pm 0.03)$ in ``Away'' (``Toward'') sector, derived in Figure \ref{fig:f3}. 
 The simulations with $B$ multiplied by the best value of $f$ also reproduce the observed 11- year variation of $D_{\rm obs}$ successfully. \par 

The ARGO-YBJ experiment reported that the observed N-S displacement is consistent with $B_{\rm IMF}$ observed at the Earth. 
In the present paper, on the other hand, we find the underestimation of $\overline{B_{\rm IMF}}$ by the PFM.
 The underestimation of the $B$ by the PFM has been recently reported also from simultaneous microwave and Extreme-Ultra Violet (EUV) observations \cite{Miyawaki2016}. 
 They found that the line-of-sight $B$ observed in the lower solar corona is $2 \sim 5.4$ times larger than $B$ calculated from the PFM and raised a question to the current-free assumption of the PFM in the photosphere and chromosphere. 
It is also reported, on the other hand, that there are significant differences between the observed photospheric magnetic fields used in the PFMs, although there is a general qualitative consensus \cite{Riley2014}. 
 The difference is arising from the difference in the observation techniques and spatial resolutions. 
 For instance, the average photospheric field strength $B_{\rm photo}$ observed by the Michelson Doppler Imager (MDI) onboard the Solar and Heliospheric Observatory (SOHO) \cite{Sherrer1995} is $1.80 \pm 0.20$ times larger than $B_{\rm photo}$ observed by the KPVT/SOLIS used in our PFM, during the recent period when both data are available for comparison (see Table 3 of \cite{Riley2014}). 
 Although this ratio is similar to $f$ obtained in this Letter, it should be noted that $\overline{B_{\rm IMF}}$ responsible for the N-S displacement is not simply proportional to $B_{\rm photo}$ in the PFM. 
 Since the $n$-th order harmonic component of the magnetic field at the radial distance $r$ is proportional to $(R_{\odot} /r)^{n+2}$ \cite{Hakamada1995}, more complex and stronger field on the photosphere represented with larger $n$ diminishes faster with increasing $r$ and only the low order harmonic components dominate the IMF at $r>R_{\rm ss}$. 
 We actually confirmed that the $B_{\rm IMF}$ at the Earth calculated by the PFM using the MDI and KPVT/SOLIS photospheric fields are quite consistent with each other. 
 The underestimation of $\overline{B_{\rm IMF}}$ by the PFM deduced from the observed Sun's shadow is, therefore, more likely due to the current-free assumption of the PFM which does not hold accurately for the plasma in the solar atmosphere.\par

	In summary, we find that the actual $\overline{B_{\rm IMF}}$ is about 1.5 times larger than the prediction by the PFM, by analyzing the angular displacement of the center of the Sun's shadow. This is unlikely due to the difference between the photospheric magnetic fields used in the PFM, but more likely due to the current-free assumption of the PFM which does not hold accurately in the plasma in the solar atmosphere. 
It is concluded that the Sun's shadow observed by the Tibet AS array, combined with other measurements, offers a powerful tool for an accurate measurement of the average solar magnetic field.

\begin{acknowledgments}
The collaborative experiment of the Tibet Air Shower Arrays has been performed under the auspices of the Ministry of Science and Technology of China (No. 2016YFE0125500) and the Ministry of Foreign Affairs of Japan. This work was supported in part by a Grant-in-Aid for Scientific Research on Priority Areas from the Ministry of Education, Culture, Sports, Science and Technology, and was supported by Grants-in-Aid for Science Research from the Japan Society for the Promotion of Science in Japan. This work is supported by the National Natural Science Foundation of China (Nos. 11533007 and 11673041) and the Chinese Academy of Sciences and the Key Laboratory of Particle Astrophysics, Institute of High Energy Physics, CAS. This work is supported by the joint research program of the Institute for Cosmic Ray Research (ICRR), the University of Tokyo. K. Kawata is supported by the Toray Science Foundation. The authors thank Dr. Xuepu Zhao of Stanford University for providing the usage of the CSSS model and Dr. K. Hakamada of Chubu University and Dr. Shiota of NICT (National Institute of Information and Communications Technology) for supplying the data calculated by the Potential Field Source Surface (PFSS) model. They also thank Dr. J\'ozsef K\'ota of the University of Arizona for his useful comments and discussions.
\end{acknowledgments}

\bibliography{PRL2017_Draft3}
\end{document}